\documentclass[a4paper,11pt]{article}
\usepackage[english]{babel}
\usepackage{xspace}
\usepackage{amsmath,amssymb}
\usepackage{mathrsfs}
\usepackage[dvips]{graphicx}
\usepackage[T1]{fontenc}
\begin{document}
\pagestyle{plain}
\newcommand{\Szero}{\ensuremath{\mathrm{S}^0}\xspace}
\newcommand{\mS}{\ensuremath{m_{\Szero}}\xspace}
\newcommand{\HSM}{\ensuremath{\mathrm{H}^0_{\mathrm{SM}}}\xspace}
\newcommand{\OPAL}{{\small OPAL}\xspace}
\newcommand{\LEP}{{\small LEP}\xspace}
\newcommand{\Zzero}{\ensuremath{\mathrm{Z}^{0}}\xspace}
\newcommand{\hzero}{\ensuremath{\mathrm{h}^{0}}\xspace}
\newcommand{\Hzero}{\ensuremath{\mathrm{H}^{0}}\xspace}
\newcommand{\dEdx}{\ensuremath{\mathrm{d}E/\mathrm{d}x}\xspace}
\newcommand{\gce}{{\small GCE}\xspace}
\newcommand{\sm}{{\small SM}\xspace}
\newcommand{\SM}{{Standard Model}\xspace}
\newcommand{\MSSM}{{\small MSSM}\xspace}
\newcommand{\mc}{Monte Carlo\xspace}
\newcommand{\MC}{Monte Carlo\xspace}
\newcommand{\klein}[1]{{\small #1}\xspace}
\newcommand{\hl}[1]{{\itshape #1}}
\newcommand{\degree}[1]{\ensuremath{\mathrm{#1}^\circ}}
\newcommand{\pa}{\ensuremath{\phi_a}\xspace}
\newcommand{\wa}{\ensuremath{\alpha}\xspace}
\newcommand{\aiso}{\ensuremath{\alpha_{\mathrm{iso}}}\xspace}
\newcommand{\minv}{\ensuremath{m_{\mathrm{inv}}}\xspace}
\newcommand{\tpmiss}{\ensuremath{\theta(\vec{p}_{\mathrm{miss}})}\xspace}
\newcommand{\pmiss}{\ensuremath{p_{\mathrm{miss}}}\xspace}
\newcommand{\nn}{\ensuremath{\nu\overline{\nu}}\xspace}
\newcommand{\mm}{\ensuremath{\mu^+\mu^-}\xspace}
\newcommand{\lplm}{\ensuremath{\mathrm{l}^+\mathrm{l}^-}\xspace}
\newcommand{\ee}{\ensuremath{\mathrm{e}^+\mathrm{e}^-}\xspace}
\newcommand{\bbar}{\ensuremath{\mathrm{b\overline{b}}}\xspace}
\newcommand{\keV}{\ensuremath{\mbox{keV}}\xspace}
\newcommand{\MeV}{\ensuremath{\mbox{MeV}}\xspace}
\newcommand{\GeV}{\ensuremath{\mbox{GeV}}\xspace}
\newcommand{\gev}{\ensuremath{\mbox{GeV}}\xspace}
\newcommand{\Evec}{\ensuremath{\vec{E}}\xspace}
\newcommand{\tbc}{(\emph{\ldots to be completed \ldots})\xspace}
\newcommand{\mrec}{\ensuremath{m_{\mathrm{r}}}\xspace}
\newcommand{\sq}{\ensuremath{k}\xspace}
\newcommand{\Nnf}{\ensuremath{\mathrm{N^{95}}}\xspace}
\newcommand{\sqnf}{\ensuremath{\mathrm{k^{95}}}\xspace}
\newcommand{\Nsm}{\ensuremath{\mathrm{N_{SM}}}\xspace}
\newcommand{\THDM}{{\small 2HDM}\xspace}
\newcommand{\SigmaZH}{\ensuremath{\sigma_{\mathrm{ZH^{SM}}}}\xspace}
\newcommand{\mH}{\ensuremath{m_{\mathrm{H}}^{\mathrm{SM}}}\xspace}
\newcommand{\mHsm}{\ensuremath{m_{\mathrm{H}}}\xspace}
\newcommand{\mh}{\ensuremath{m_{\mathrm{h^0}}}\xspace}
\newcommand{\mhwo}{\ensuremath{m}\xspace}
\newcommand{\mhi}{\ensuremath{m_{\mathrm{h^0_i}}}\xspace}
\newcommand{\hi}{\ensuremath{\mathrm{h^0_i}}\xspace}
\newcommand{\Ki}{\ensuremath{K_i}\xspace}
\newcommand{\SigmaZh}{\ensuremath{\sigma_{\mathrm{Zh}}}\xspace}
\newcommand{\mA}{\ensuremath{m_{\mathrm{A}}}\xspace}
\newcommand{\mB}{\ensuremath{m_{\mathrm{B}}}\xspace}
\newcommand{\mC}{\ensuremath{m_{\mathrm{C}}}\xspace}
\newcommand{\intKdm}{\ensuremath{\int \tilde{K}\mathrm{d}m}\xspace}
\newcommand{\Ktilde}{\ensuremath{\tilde{K}}\xspace}
\newcommand{\err}[3]{\,\ensuremath{\mathrm{#1}\pm\mathrm{#2\,(stat.)}\pm\mathrm{#3\,(syst.)}}\xspace}
\newcommand{\erro}[2]{\,\ensuremath{\pm\mathrm{#1}\pm\mathrm{#2}}\xspace}
\newcommand{\intk}{\ensuremath{\int_{\mA}^{\mB} \tilde{K}\,\mathrm{d}m}\xspace}
\newcommand{\mathd}{\ensuremath{\mathrm{d}}}
\newcommand{\deltam}{\ensuremath{\Delta m}}
\newcommand{\rb}[1]{\raisebox{-2ex}{#1}}
\newcommand{\pb}{\ensuremath{\mathrm{pb}^{-1}}}
\newcommand{\Ztoee}{\ensuremath{\Zzero\to\ee}}
\newcommand{\Ztomm}{\ensuremath{\Zzero\to\mm}}
\newcommand{\vnr}[1]{\vphantom{\rule{0mm}{#1}}\xspace}
\newcommand{\rr}{\raggedright\small}
\newcommand{\prelim}{\large\textbf{Preliminary}\xspace}
\newcommand{\G}{\mbox{$\mathrm{GeV}$}}
\newcommand{\sqrts}{\mbox{$\sqrt {s}$}}
\newcommand{\mZ}{$m_{\mathrm{Z}}$\xspace}
\newcommand{\bb}{\mbox{$\mathrm{b}\bar{\mathrm{b}}$}}
\newcommand{\etal}{\mbox{\it et al.}}
\begin{titlepage}
\thispagestyle{empty}
\vspace*{0.075\textheight}

\begin{center}
  \textbf{\Huge Decay-mode independent searches\\[1ex] 
    for new scalar bosons with OPAL}\\[6ex]
  \renewcommand{\thefootnote}{\fnsymbol{footnote}}
  {\Large Jochen Cammin\footnote{Physikalisches Institut, Nussallee~12,
    D-53115 Bonn, Germany, email: cammin@physik.uni-bonn.de}\\[1ex]
    \itshape{Physikalisches Institut der Universität Bonn, Germany}}
  \setcounter{footnote}{0}
  \bigskip\bigskip
  
  { Talk given at the 10th International Conference on Supersymmetry
    and Unification of Fundamental Interactions, DESY Hamburg, Germany,
    June~2002} \bigskip\bigskip\bigskip\bigskip

\end{center}
\begin{abstract}
  \noindent Topological searches for neutral scalar bosons
  \Szero produced in association with a \Zzero boson via the Bjorken
  process $\ee\to\Szero{}\Zzero$ at centre-of-mass energies of 91~\GeV
  and 183--209~\GeV are described.  These searches are based on
  studies of the recoil mass spectrum of $\Zzero\to\ee$ and $\mu^+
  \mu^-$ events and on a search for $\Szero\Zzero$ with $\Zzero \to
  \nu\bar{\nu}$ and $\Szero \to \ee$ or photons.  They cover the
  decays of the \Szero into an arbitrary combination of hadrons,
  leptons, photons and invisible particles as well as the possibility
  that it might be stable.
  
  No indication for a signal is found in the data and upper limits on
  the cross section of the Bjorken process are calculated.
  Cross-section limits are given in terms of a scale factor \sq with
  respect to the \SM cross section for the Higgs-strahlung process
  $\ee\to\HSM\Zzero$.
  
  These results can be interpreted in general scenarios independently
  of the decay modes of the \Szero. The examples considered here are
  the production of a single new scalar particle with a decay width
  smaller than the detector mass resolution, and for the first time,
  two scenarios with continuous mass distributions, due to a single
  very broad state or several states close in mass.
\end{abstract}
\end{titlepage}


  
  

\section{Introduction}
Searches for new neutral CP even scalar bosons \Szero with the
\OPAL~\cite{c:detector} detector at \LEP are described. The new bosons
are assumed to be produced in association with a \Zzero boson via the
Bjorken process $\ee \to \mathrm{\Szero{}\Zzero}$. 

The analyses are topological searches and are based on studies of the
recoil mass spectrum in $\Zzero\to\ee$ and $\mu^+ \mu^-$ events and on
a subsequent search for $\Szero\Zzero$ events with $\Szero \to \ee$ or photons
and $\Zzero \to \nu\bar{\nu}$.  They are sensitive to all decays of
\Szero into an arbitrary combination of hadrons, leptons, photons and
invisible particles, and to the case of a long-lived \Szero leaving
the detector without interaction. Hence they are decay-mode
independent and result in absolute mass limits. The analyses are applied to 
\LEP~1 \Zzero on-peak data (115.4~pb$^{-1}$ at $\sqrt{s}=91.2~\GeV$)
and to 662.4~pb$^{-1}$ of \LEP~2 data collected at centre-of-mass
energies in the range of 183 to 209~\GeV. 
The results are presented in terms of limits on the scaling factor
\sq, which relates the \Szero{}\Zzero production cross section to the
Standard Model (\klein{SM}) cross section for the Higgs-strahlung
process:
\begin{equation}\label{e:sq_def}
  \sigma_{\mathrm{\Szero\Zzero}} = \sq\cdot
  \sigma_{\mathrm{\HSM\Zzero}}(m_{\mathrm{\HSM}}=m_{\mathrm{\Szero}}),
\end{equation} 
where it is assumed that \sq does not depend on the centre-of-mass
energy for any given mass $m_{\mathrm{\Szero}}$. Since the analysis is
independent of the decay mode of the \Szero, these limits can be
interpreted in any scenario beyond the Standard Model. 
The most general case is to provide upper limits on the
  cross section or scaling factor \sq for a single narrow new scalar boson
  independent of its couplings to other particles. 
  In a more specific interpretation, assuming the \Szero{}\Zzero
  production cross section to be identical to the Standard Model Higgs
  boson one, the limit on \sq can be translated into a lower limit on
  the Higgs boson mass\footnote{Dedicated searches for the Standard
    Model Higgs boson by the four \LEP experiments, exploiting the
    prediction for its decay modes, have ruled out masses of up to
    114.4~\GeV \cite{c:LEP_Higgs_limit}.}.
%
For the first time we give limits not only for a single mass
  peak with small width, but also for a continuous distribution of the
  signal in a wide mass range. Such continua appear in several
  recently proposed models  
  which are introduced in the next sections.

\subsection{Continuous Higgs scenarios}
\subsubsection*{The Uniform Higgs scenario}\label{worst-case}
This model, as described in Ref.~\cite{c:gunion}, assumes a broad
enhancement of the signal over the background expectation in the $M_{\mathrm{X}}$
mass distribution for the process $\ee\to \Zzero\mathrm{X}$.  This
enhancement is due to numerous additional neutral Higgs bosons \hi
with masses $m_{\mathrm{A}} \le m(\hi) \le m_{\mathrm{B}}$, where
$m_{\mathrm{A}}$ and $m_{\mathrm{B}}$ indicate the lower and upper
bound of the mass spectrum.  The squared coupling, $g^2$, of the Higgs states \hi
to the \Zzero is modified by a factor $k_i$ compared to the \SM
\Hzero{}\Zzero coupling: $g^2_{\Zzero\hi} = k_i\cdot
g^2_{\Zzero\hzero_{\mathrm{SM}}}(m_{\hi}=m_{\hzero_{\mathrm{SM}}})$.
  
If the Higgs states are assumed to be closer in mass than the
experimental mass resolution, then there is no means to distinguish
between separate $k_i$.  In this case the Higgs states and their
reduction factors $k_i$ can be combined into a coupling density
function, $\Ktilde(m) = \mathrm{d}k/\mathrm{d}m$.  The model obeys two
sum rules which in the limit of unresolved mass peaks can be expressed
as integrals over this coupling density function:
\begin{eqnarray} 
  \int\limits_{0}^{\infty} \mathd\mhwo\; \tilde{K}(\mhwo) =  1 \label{eq:sumrule1} 
  &\quad\mbox{and}\quad&
  \int\limits_{0}^{\infty} \mathd\mhwo\; \tilde{K}(\mhwo){\mhwo}^2 \le  m^2_\mathrm{C}, 
  \label{eq:sumrule2} 
\end{eqnarray}
where $\Ktilde(m) \ge 0$ and \mC is a perturbative mass scale of the
order of 200~\GeV.  The value of \mC is model dependent and can be
derived by requiring that there is no Landau pole up to a scale
$\Lambda$ where new physics occurs \cite{c:gunion}.  If neither a
continuous nor a local excess is found in the data,
Equation~\ref{eq:sumrule1} can be used to place constraints on the
coupling density function $\Ktilde(m)$. For example, if $\Ktilde(m)$
is assumed to be constant over the interval [\mA, \mB{}] and zero
elsewhere,
then certain choices for the interval [\mA, \mB{}] can be excluded. From
this and from Equation~\ref{eq:sumrule2} lower limits on the mass
scale \mC can be derived.

\subsubsection*{The Stealthy Higgs scenario}
This scenario predicts the existence of additional SU(3)$_{\mathrm{C}}
\times$SU(2)$_{\mathrm{L}}\times$U(1)$_{\mathrm{Y}}$ singlet fields
(phions), which would not interact via the strong or electro-weak
forces, thus coupling only to the Higgs boson with strength $\omega$ 
\cite{c:stealthy_higgs}.
Therefore these singlets would reveal their existence only in the
Higgs sector by offering invisible decay modes to the Higgs boson. The
width of the Higgs resonance can become large if the number of such
singlets, $N$, or the coupling $\omega$ is large, thus yielding a
broad spectrum in the mass recoiling against the reconstructed \Zzero.
The interaction term between the Higgs and the additional phions in
the Lagrangian is given by
$
  \mathscr{L}_{\mathrm{interaction}} = 
    -\frac{\omega}{2\sqrt{N}}
    \vec{\varphi}^2\phi^\dagger\phi,
$
where $\phi$ is the Standard Model Higgs doublet, $\omega$ is the
coupling constant, and $\vec{\varphi}$ is the vector of the new
phions. An analytic expression for the Higgs width can be found in the
limit $N\to\infty$:
\begin{equation}\label{eq:higgs_width}
  \Gamma_{\mathrm{H}}(\mHsm) = \Gamma_{\mathrm{SM}}(\mHsm) + \frac{\omega^2 v^2}
                  {32\, \pi\, \mHsm},
\end{equation}
where $v$ is the vacuum expectation value of the Higgs field.  This
expression results when setting other model parameters to zero,
including the mass of the phions \cite{c:stealthy_higgs}.  

In section~\ref{s:stealthy} we derive limits on the Stealthy Higgs
model. By simulating signal spectra for different Higgs widths
$\Gamma_{\mathrm{H}}$ we constrain the $\omega$-$m_\mathrm{H}$ plane
in the large $N$ limit.

\section{Decay-mode independent searches for 
\boldmath e$^+$e$^-\to$S$^0$Z$^0$\unboldmath}
\label{s:decay_independent_searches}
The event selection is intended to be efficient for the complete
spectrum of possible \Szero decay modes.  As a consequence it is
necessary to consider a large variety of 2-fermion, 4-fermion, and
2-photon background processes.
Suppression of the background is performed using the smallest amount
of information possible for a particular decay of the \Szero.  The
decays of the \Zzero into electrons and muons are the channels with
highest purity, and therefore these are used in this analysis. 
The signal process can be tagged by identifying events with an
acoplanar, high momentum electron or muon pair.

Different kinematics of the processes in the \LEP~1 and the \LEP~2
analysis lead to different strategies for rejecting the background.
At \LEP~2 the invariant mass of the two final-state leptons in the
signal channels is usually consistent with the \Zzero mass, while this
is not true for a large part of the background. Therefore a cut on the
invariant mass rejects a large amount of background. Remaining
two-fermion background from radiative processes can partially be
removed by using a photon veto without losing efficiency for photonic
decays of the \Szero.  In the \LEP~1 analysis the invariant mass of
the lepton pair cannot be constrained. Therefore, stronger selection
cuts have to be applied to suppress the background, resulting in an
insensitivity to the decays $\Szero \to $ photons and $\Szero \to \ee$
at low masses. These decay modes are recovered in a search dedicated
to $\ee\to \Szero\Zzero$ with $\Zzero \to \nu\bar{\nu}$ and $\Szero
\to$ photons (or photons plus invisible particles) or \ee. Details of
the selection procedure can be found in~\cite{c:dmi}.

\begin{table}
  \centering
  \small
\begin{tabular}{|c|c|r||c|c|c||r|}\hline
\rb{$\sqrt{s}$~(GeV)} &  
\rb{Data}             & 
\multicolumn{1}{c||}{\raisebox{-1ex}{Total}}&
\rb{2-fermion}        & 
\rb{4-fermion}        &  
\rb{2-photon}         &  
\multicolumn{1}{c|}{\rb{Signal}} \\
& & \multicolumn{1}{c||}{\raisebox{1ex}{bkg.}} & & & &  \multicolumn{1}{c|}{\raisebox{1mm}{($m_{\Szero}$=30~GeV)}} \\ 
\hline\hline
\multicolumn{1}{|c}{ }& \multicolumn{6}{c|}{Electron channel}\\
\hline
91.2 &  45 & 55.2\erro{3.0}{3.0} & 20.5 & 34.4 & 0.3  & 15.61\erro{0.31}{0.47}\\ \hline
183--209
     &  54 & 46.9\erro{0.6}{3.5} & 12.8 & 33.7 & 0.4  &  7.97\erro{0.06}{0.25} \\
\hline\hline
\multicolumn{1}{|c}{ }& \multicolumn{6}{c|}{Muon channel}\\
\hline
91.2 &  66 & 53.6\erro{2.7}{2.1} & 17.0 & 35.4 & 1.2  & 21.55\erro{0.45}{0.69}\\ \hline
183--209
     &  43 & 51.6\erro{0.3}{2.5} &12.2  & 38.6 & 0.8  &  9.43\erro{0.06}{0.37} \\
\hline
\end{tabular}

\caption{\label{t:d_b_error} 
  Selected data events, background Monte Carlo and signal 
  expectation for a 30~GeV \SM  Higgs boson. The first
  error is statistical and the second error is systematic. 
  }
\end{table}

\section{Results}\label{s:results}
\begin{figure}
  \centering
  \includegraphics[width=0.5\linewidth]{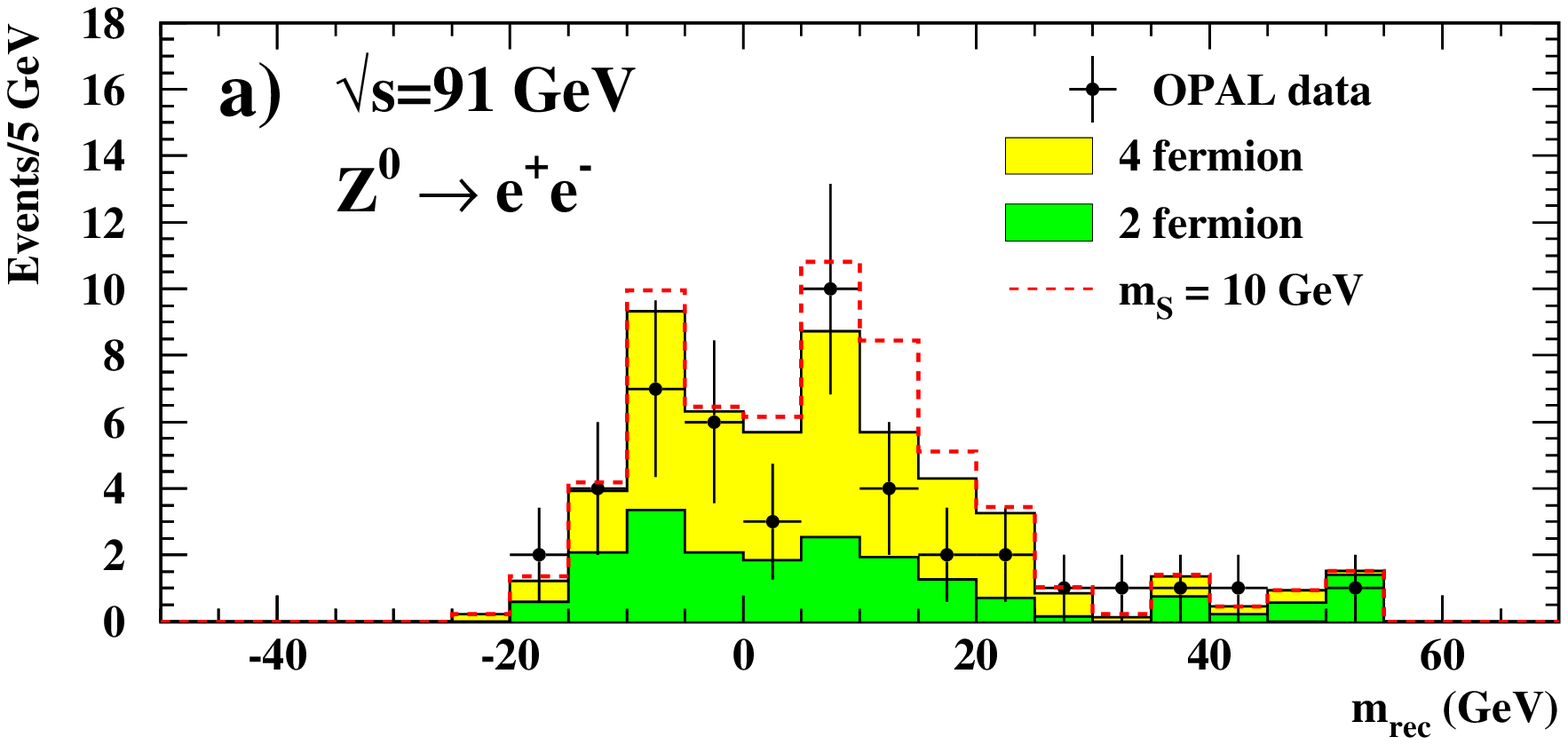}
  \includegraphics[width=0.5\linewidth]{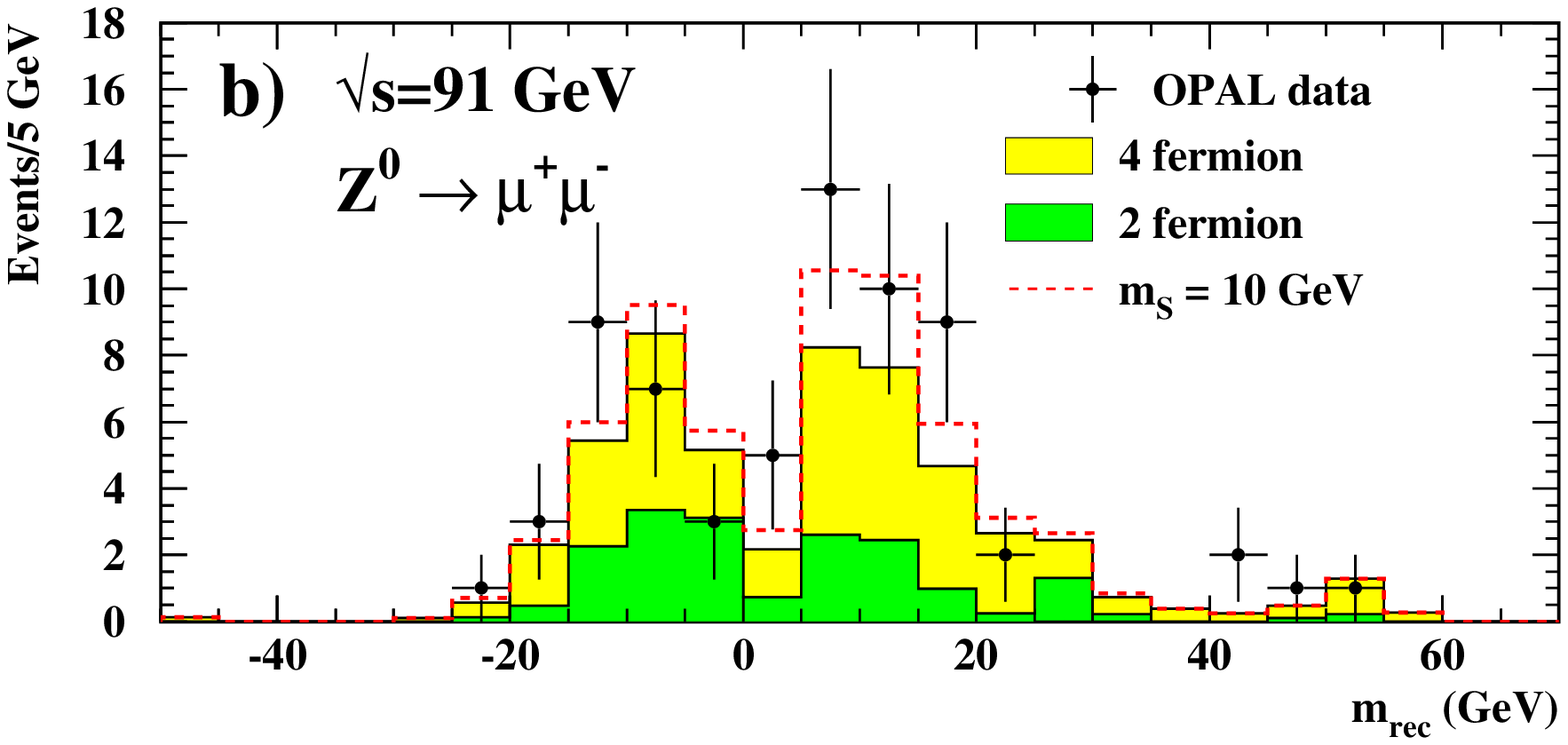}
  \caption{\label{f:summass_LEP1} 
    The recoil mass spectra from $\sqrt{s}=$91.2~GeV a) for the decays
    $\Zzero \to \ee$ and b) for $\Zzero \to \mm$. \klein{OPAL} data are
    indicated by dots with error bars (statistical error), the
    four-fermion background by the light grey histograms and the
    two-fermion background by the dark grey histograms. The dashed lines
    for the signal distributions are plotted on top of the background
    distributions with normalisation corresponding to the cross section
    excluded at 95\% confidence level from the combination of both
    channels.
  }
  \vspace*{0.1\textheight}

  \includegraphics[width=0.5\linewidth]{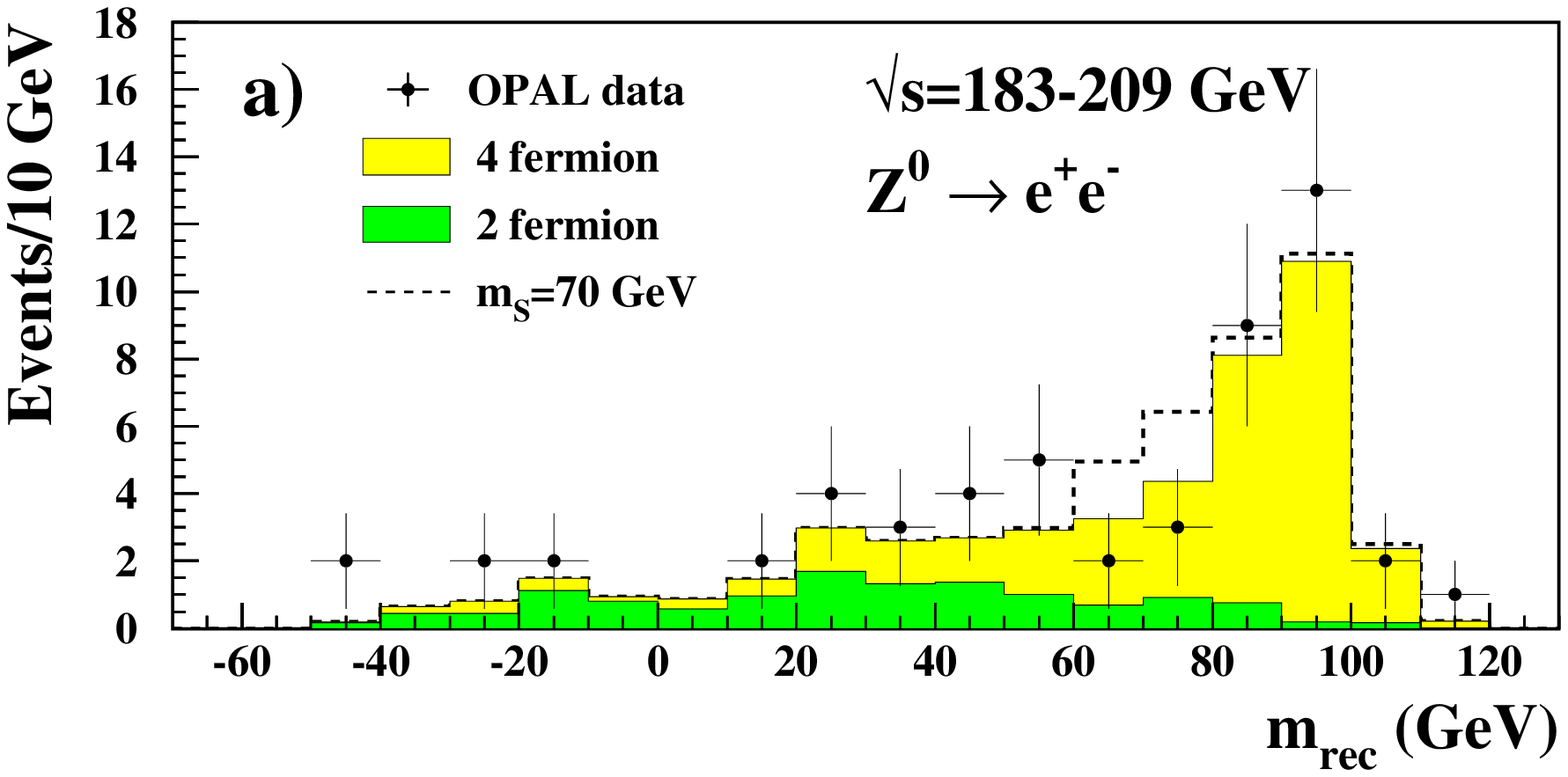}
  \includegraphics[width=0.5\linewidth]{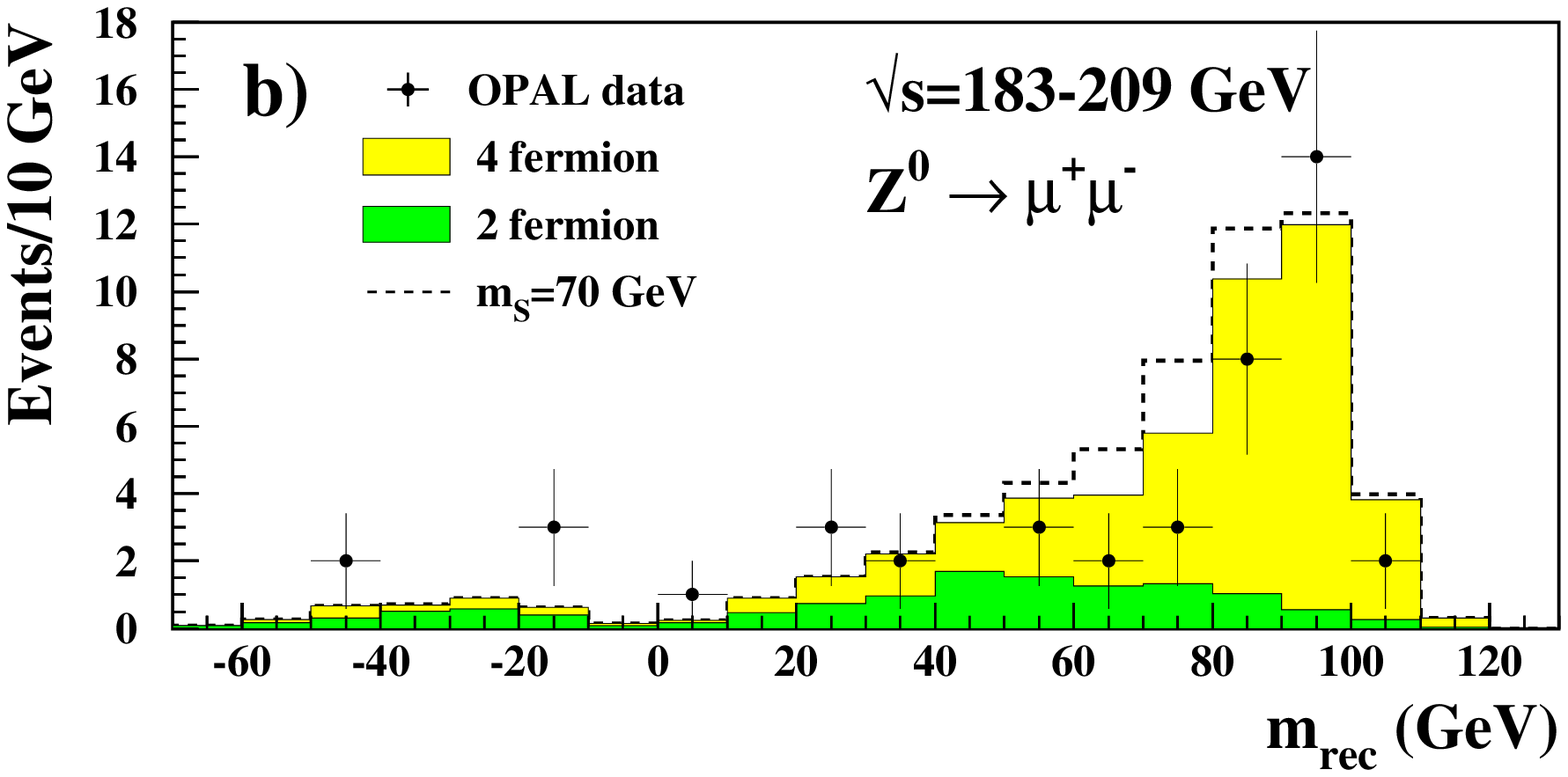}
  \caption{\label{f:summass_LEP2} 
    The recoil mass spectrum from 183--209~GeV a) for the decays $\Zzero
    \to \ee$ and b) for $\Zzero \to \mm$ (lower plot). \klein{OPAL} data
    are indicated by dots with error bars (statistical error), the
    four-fermion background by the light grey histograms and the
    two-fermion background by the medium grey histograms. The dashed
    lines for the signal distributions are plotted on top of the
    background distributions with normalisation corresponding to the
    excluded cross section from the combination of both channels.
  }
\end{figure}

The results of the decay-mode independent searches are summarised in
Table~\ref{t:d_b_error}.
The total number of observed candidates from all lepton channels
combined is 208, while the \SM background expectation amounts to
$207.3\pm 4.1(\mathrm{stat.}) \pm 11.1(\mathrm{syst.})$. The
efficiency for the signal process is around 30\,\% for most of the
mass range. Figure~\ref{f:summass_LEP1} and \ref{f:summass_LEP2} show
the recoil mass spectra for the lepton channels at \LEP~1 and \LEP~2.
As no excess over the expected background is observed in the data,
limits on the cross section for the Bjorken process
$\mathrm{e}^+\mathrm{e}^-\to$ \Szero{}\Zzero\ are calculated.
  
The limits are presented in terms of a scale factor \sq, which relates
the cross section for \Szero{}\Zzero to the \SM\ one for the
Higgs-strahlung process $\mathrm{e}^+\mathrm{e}^-\to$ \HSM{}\Zzero as
defined in Equation~\ref{e:sq_def}. The 95\% confidence level upper
bound on $\sq$ is obtained from a test statistic, ${\cal
  L}_{s+b}/{\cal L}_{b}$, for the signal and the signal+backgrund
likelihood, by using the recoil mass distributions of the data, the
background and the signal (applying the Higgs decay with the smallest
efficiency) and the weighted event-counting method described
in~\cite{c:mssmpaper172}. The systematic uncertainties are included as
described in~\cite{c:cousins}. To give a conservative limit only the
lepton channels are used in the limit calculation, because the channel
with $\Zzero\to\nn$ is complementary to the lepton channels in the
\LEP~1 search, and it has a much higher sensitivity.
  
The limits are given for three different scenarios:
1. Production of a single new scalar {\boldmath
  S$^0$\unboldmath}
\quad 2. The Uniform Higgs scenario
\quad 3. The Stealthy Higgs scenario.

\subsection{Production of a single new scalar \boldmath S$^0$\unboldmath}
\label{general}
In the most general interpretation of the results, a cross-section
limit is set on the production of a new neutral scalar boson \Szero in
association with a \Zzero boson. 

\begin{figure}
  \centering
  \includegraphics[width=0.55\textwidth]{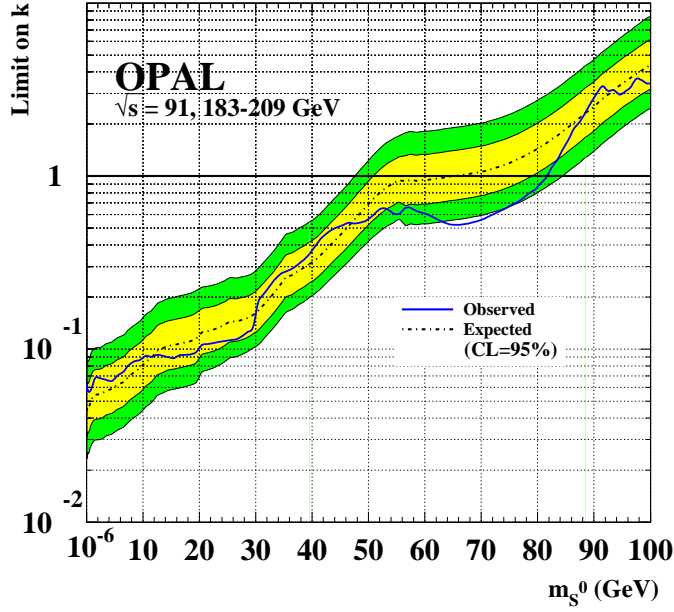}
\caption{
  The upper limit on the scale factor \sq on the cross section for the
  production of a new scalar boson in the Higgs-strahlung process
  (solid line). The dot-dashed line represents the expected median for
  background-only experiments.  Both limits are calculated at the
  95\,\% confidence level.  The dark (light) shaded bands indicate the
  68\% (95\%) probability intervals centred on the median expected
  values.  For masses $\mS \lesssim 1~\GeV$ the limits are constant.
  The lowest signal mass tested is $10^{-6}~\GeV$.  
  }
\label{f:di_limits}
\end{figure}

In Figure~\ref{f:di_limits} the limits obtained for scalar
masses down to the lowest generated signal mass of 1~\keV are shown. They are
valid for the decays of the \Szero into hadrons, leptons, photons and
invisible particles (which may decay inside the detector) as well as
for the case in which the \Szero has a sufficiently long lifetime to
escape the detector without interacting or decaying.  
The observed limits are given by the solid line, while the expected
sensitivity, determined from a large number of Monte Carlo experiments
with the background-only hypothesis, is indicated by the dotted line. The shaded
bands indicate the one and two sigma deviations from the expected
sensitivity. 
The existence of a Higgs boson produced at the SM rate can
be excluded up to 81~\GeV even from decay-mode independent searches.
For masses of the new scalar particle well below the width of the
\Zzero, \emph{i.e.} $\mS \lesssim 1~\GeV$, the obtained limits remain
constant at the level of $\sq^{95}_{\mathrm{obs.}} = 0.067$, and
$\sq^{95}_{\mathrm{exp.}} = 0.051$.

\subsection{Limits on signal mass continua}
\subsubsection{The Uniform Higgs scenario\label{s:uniform-higgs}}
\begin{figure}
  \centering
  \includegraphics[width=0.55\textwidth]{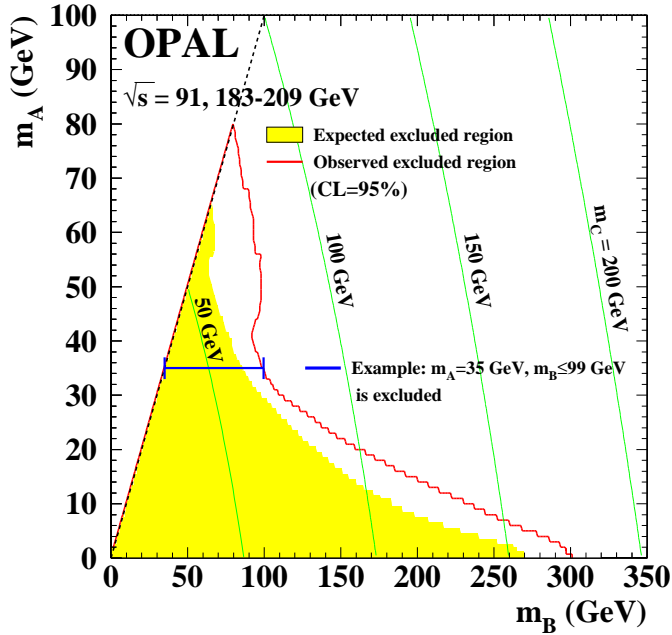}
\vspace*{-2ex}

\caption{
  Exclusion limits for the Uniform Higgs scenario at the 95\,\%
  confidence level. All mass intervals (\mA,\,\mB) within the area
  bordered by the dark line are excluded from the data. The shaded
  area marks the mass points which are expected to be excluded if
  there were only background.  The light grey curves indicate isolines
  for several values of \mC. All intervals $(\mA,\mB)$ to the right of
  each isoline are theoretically disallowed from
  Equation~\ref{eq:sumrule2}. By definition, only intervals
  $(\mA,\mB)$ right to the dashed diagonal line are valid, \emph{i.e.}
  $\mA \le \mB$.
  \label{fig:excluded_continuum_higgs}
  }
\end{figure}

Limits are obtained for the Uniform Higgs scenario where $\Ktilde =
\mathrm{constant}$ over the interval [\mA, \mB{}] and zero elsewhere.
Both the lower mass bound \mA and the upper bound \mB are varied
between 1~\GeV and 350~\GeV (with the constraint $\mA\le\mB$).  
An upper limit is set on the first integral in Equation~\ref{eq:sumrule1}.
Figure~\ref{fig:excluded_continuum_higgs} shows the mass points
(\mA,\mB) for which the obtained 95\,\%~CL limit on
$\int\mathrm{d}m\,\Ktilde$ is less than one.  These are the signal
mass ranges $\mA \le \mhi \le \mB$ which can be excluded assuming a
constant \Ktilde.

The horizontal line illustrates an example for excluded mass ranges:
The line starts on the diagonal at $\mA = \mB = 35~\GeV$ and ends at
$\mB = 99~\GeV$. This value of \mB is the highest upper mass bound
which can be excluded for this value of \mA.  All mass ranges with an
upper bound \mB below 99~\GeV are also excluded for $\mA = 35~\GeV$.
The highest excluded value of \mB ($\mB = 301~\GeV)$ is achieved for
$\mA$ set to 0~\GeV.

\begin{figure}
  \centering
  \includegraphics[width=0.55\textwidth]{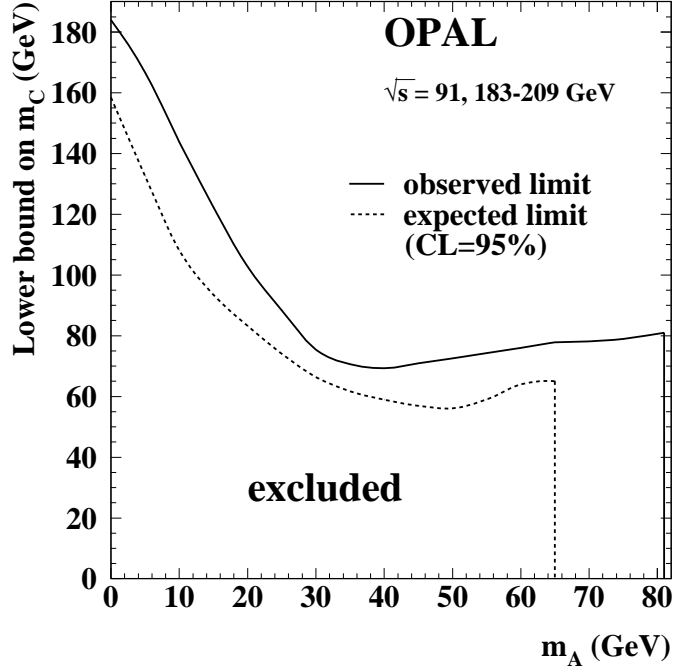}
\vspace*{-2ex}

\caption{
  Exclusion limits on the perturbative mass scale \mC for constant
  \Ktilde. The solid line represents the limits obtained from the
  data, and the dotted line shows the expected limit if there were
  only background. Values for \mC below the lines are excluded by this
  analysis at the 95\,\% confidence level.  
  }
  \label{fig:excluded_mC}
\end{figure}

Using the maximal exluded mass ranges and the two sum rules from
section~\ref{worst-case} for constant \Ktilde, lower limits on the
perturbative mass scale \mC can be derived, as shown in
Figure~\ref{fig:excluded_mC}.
It is also possible to set limits on a non-constant coupling density.
Details can be found in Ref.~\cite{c:dmi}.

\subsubsection{The Stealthy Higgs scenario}
\label{s:stealthy}
To set limits on the Stealthy Higgs scenario we have simulated the
spectrum of a Higgs boson with a width according to
Equation~\ref{eq:higgs_width} and the cross section from
Ref.~\cite{c:stealthy_higgs}.

\begin{figure}
  \centering
  \includegraphics[width=0.55\textwidth]{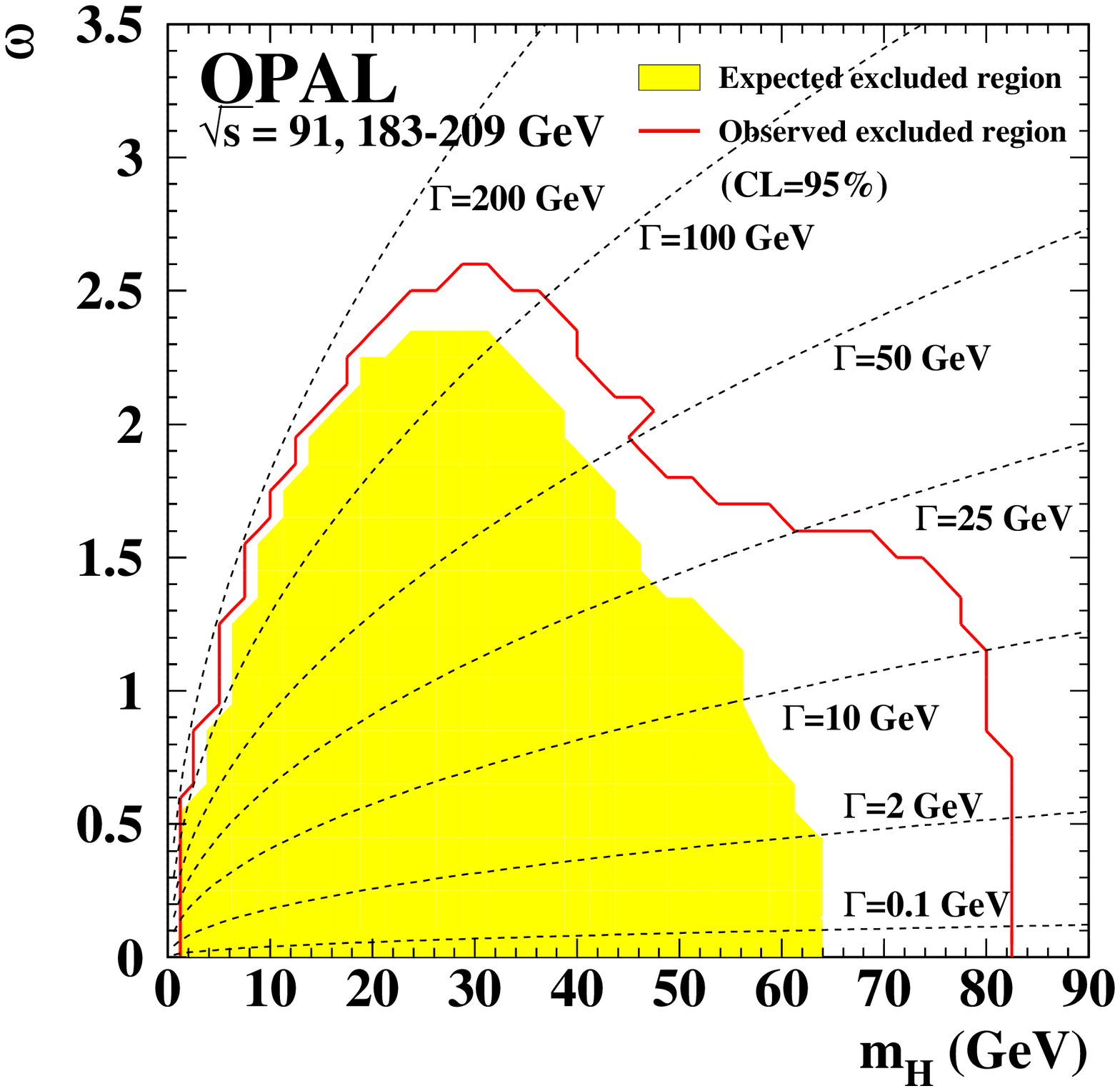}
\caption{
  Excluded parameter regions for the Stealthy Higgs scenario at the
  95\,\% confidence level. The solid line marks the region which is
  excluded from the data. The shaded area marks the region which would
  be excluded if there were only background.  The dashed lines
  indicate the Higgs width depending on $m_{\mathrm{H}}$ and $\omega$.
  \label{fig:excluded_hidden_higgs}
  }
\end{figure}

The excluded regions in the $\omega$-$m_{\mathrm{H}}$ parameter space
are shown in Figure~\ref{fig:excluded_hidden_higgs}. To illustrate the
Higgs width according to Equation~\ref{eq:higgs_width}, for a given
mass $m_{\mathrm{H}}$ and coupling $\omega$ `isolines' for some sample
widths are added to the plot. 
The maximal excluded region of the coupling $\omega$ is achieved for
masses around 30~\GeV, where $\omega$ can be excluded up to $\omega =
2.7$.  For lower masses the sensitivity drops due to the rapidly
increasing width of the Higgs boson, and for higher masses due to the
decreasing signal cross section.

\section{Conclusions}
Decay-mode independent searches for new neutral scalar bosons \Szero
decaying to hadrons of any flavour, to leptons, photons invisible
particles and other modes have been performed based on the data
collected at $\sqrt{s}$ = \mZ and 183 to 209~\GeV by studying
\Szero{}\Zzero production in the channels with $\Zzero\to\ee, \mu^+
\mu^-$ and the channel where the \Zzero decays into $\nn$ and the
\Szero into photons or \ee. No excess of candidates in the data over
the expected Standard Model background has been observed.  Upper
limits on the production cross section for associated production of
\Szero and \Zzero, with arbitrary \Szero decay modes, were set at the
95\,\% confidence level.  Upper limits in units of the \SM
Higgs-strahlung cross section of $\sq < 1$ for $\mS < 81~\GeV$ were
obtained.  In further interpretations, limits on broad continuous
signal mass shapes to which previous analyses at \LEP had no or only
little sensitivity were set for the first time.  Two general scenarios
in the Higgs sector were investigated: A uniform scenario, when the
signal arises from many unresolved Higgs bosons, and a Stealthy Higgs
model, when the Higgs resonance width is large due to large
Higgs-phion couplings.
\bigskip
\bigskip

\noindent This work was supported by the German Ministerium für Bildung,
Wissenschaft, Forschung und Technologie (BMBF) under contract
no.~05HA8PD1.

\begin{thebibliography}{99}
\bibitem{c:detector}
The OPAL Collaboration, K.~Ahmet \etal, Nucl. Instr. and Meth. 
{\bf A305} (1991) 275;\\
B.~E.~Anderson \etal, IEEE Trans. on Nucl. Science 41 (1994) 845;\\
S. Anderson \etal, Nucl. Instr. and Meth. {\bf A403} (1998) 326;\\
G.~Aguillion \etal, Nucl. Instr. and Meth. {\bf A417} (1998) 266.
\bibitem{c:LEP_Higgs_limit}
  ALEPH, DELPHI, L3 and OPAL Collaborations, The LEP working group for
  Higgs boson searches, LHWG Note/2002-01,\\
  C.~Mariotti, talk at the $31^{\mathrm{st}}$
  International Conference on High Energy Physics, ICHEP2002, Amsterdam.
\bibitem{c:gunion}
  J.R.~Espinosa and J.F.~Gunion,
  Phys.\ Rev.\ Lett.\  {\bf 82} (1999) 1084.
\bibitem{c:stealthy_higgs}
  T.~Binoth and J.J.~van der Bij,
  Z.\ Phys.\ {\bf C75} (1997) 17;\\
  T.~Binoth and J.J.~van der Bij, 
  hep-ph/9908256.
\bibitem{c:dmi}
  The OPAL Collaboration, G.~Abbiendi \etal,
  hep-ex/0206022,
  submitted to Eur.~Phys.~J.~\textbf{C}.
\bibitem{c:mssmpaper172} 
  The OPAL Collaboration, K.~Ackerstaff \etal, 
  Eur. Phys. J. {\bf C5} (1998) 19.
\bibitem{c:cousins}
  R.~D.~Cousins and V.~L.~Highland,
  Nucl.\ Instrum.\ Meth. {\bf A320} (1992) 331.
\end{thebibliography}
\end{document}